\documentclass[a4paper,UKenglish]{scrartcl}

\usepackage{amssymb}
\usepackage{amsmath}
\usepackage{amsfonts}
\usepackage{mathtools}

\usepackage{url}
\usepackage{hyperref}
\usepackage{cleveref}

\usepackage[dvipsnames]{xcolor}
\usepackage{xcolor-solarized}

\usepackage{subcaption}

\usepackage{threeparttable,booktabs}
\usepackage{tabularx}
\usepackage{makecell}

\usepackage[group-minimum-digits=4,group-separator={,},binary-units]{siunitx}
\DeclareSIUnit\persecond{/s}

\newcommand{\Oh}[1]{\mathcal{O}\left({#1}\right)}

\newcommand{\Th}[1]{\Theta\left({#1}\right)}

\newcommand{\setN}{\mathbb{N}}

\newcommand{\lce}[0]{\text{lce}}
\newcommand{\rmq}[0]{\text{rmq}}
\newcommand{\qpred}[0]{\text{pred}}
\newcommand{\qsucc}[0]{\text{succ}}

\usepackage{microtype}

\usepackage{amsthm}
\newtheorem{lemma}{Lemma}
\theoremstyle{definition}
\newtheorem{example}{Example}

\usepackage[linesnumbered,noend]{algorithm2e}
\SetFuncSty{}
\SetArgSty{}
\SetKwInOut{Input}{Input}
\SetKwInOut{Output}{Output}
\SetKw{Break}{break}
\SetKw{Continue}{continue}
\SetKw{Or}{or}
\SetKw{And}{and}
\SetKwProg{Fn}{Function}{:}{}

\usepackage{tikz,pgfplots}

\usepgfplotslibrary{groupplots}

\usetikzlibrary{positioning}
\usetikzlibrary{decorations.pathreplacing,calligraphy}
\usetikzlibrary{arrows.meta}
\usetikzlibrary{shapes.geometric}
\usetikzlibrary{matrix}

\tikzset{
    scriptsize/.style={
        font=\scriptsize
    }
}

\pgfplotscreateplotcyclelist{my-colors}{%
    solarized-red, every mark/.append style={solid,scale=0.67}, mark=square* \\
    solarized-blue, every mark/.append style={solid,scale=1}, mark=triangle* \\
    solarized-yellow, every mark/.append style={solid,scale=1}, mark=* \\
    solarized-orange, every mark/.append style={solid,scale=1}, mark=diamond* \\
    solarized-violet, every mark/.append style={solid,scale=1,semithick}, mark=x \\
    solarized-green, every mark/.append style={solid,scale=1,semithick}, mark=star \\
    solarized-cyan, every mark/.append style={solid,scale=1,semithick}, mark=asterisk \\
    solarized-magenta, every mark/.append style={solid,scale=1,semithick}, mark=Mercedes star \\
    solarized-orange, every mark/.append style={solid,scale=1,semithick}, mark=Mercedes star flipped \\
}

\pgfplotsset{
    major grid style = { thin, dotted, color = black!50 },
    minor grid style = { thin, dotted, color = black!50 },
    grid,
    every tick label/.append style={font=\tiny},
    every axis label/.style={font=\scriptsize},
    every axis/.append style={
        cycle list name=my-colors,
    },
    xlabel near ticks,
    ylabel near ticks,
    group3/.style={
    
    }
}

\newcommand{\twodots}{..}

\title{Computing the LZ-End parsing: Easy to implement and practically efficient}
\author{Patrick Dinklage}

\begin{document}

\maketitle

\begin{abstract}
The LZ-End parsing [Kreft \& Navarro,~2011] of an input string yields compression competitive with the popular Lempel-Ziv 77 scheme, but also allows for efficient random access.
Kempa and Kosolobov showed that the parsing can be computed in time and space linear in the input length [Kempa \& Kosolobov,~2017], however, the corresponding algorithm is hardly practical.
We put the spotlight on their suboptimal algorithm that computes the parsing in time $\Oh{n \lg\lg n}$.
It requires a comparatively small toolset and is therefore easy to implement, but at the same time very efficient in practice.
We give a detailed and simplified description with a full listing that incorporates undocumented tricks from the original implementation, but also uses lazy evaluation to reduce the workload in practice and requires less working memory by removing a level of indirection.
We legitimize our algorithm in a brief benchmark, obtaining the parsing faster than the state of the art.
\end{abstract}

\section{Introduction}
\label{sect:intro}

LZ-End is a dictionary compression scheme proposed by Kreft~and~Navarro~\cite{DBLP:conf/dcc/KreftN10} that combines useful properties of the famous schemes by Lempel~and~Ziv, \emph{LZ77}~\cite{DBLP:journals/tit/ZivL77} and \emph{LZ78}~\cite{DBLP:journals/tit/ZivL78}.
Like LZ78 and unlike LZ77, it allows for efficient random access on the compressed string.
Like LZ77 and unlike LZ78, phrases in the LZ-End parsing do not extend one another by strictly only one character, but can capture longer repetitions to result in a smaller parsing.

We focus on the computation of the LZ-End parsing of a string $S$ of length $n$.
In the inaugural work, Kreft~and~Navarro make use of backward searches on the Burrows-Wheeler transform of the reverse input $\overleftarrow{S}$, i.e., they use the FM-index~\cite{DBLP:journals/jacm/FerraginaM05}.
With this, they achieve a running time of $\Oh{nh(\lg\sigma + \lg\lg n)}$, where $h$ is the length of the longest LZ-End phrase and $\sigma = |\Sigma|$ the size of the input alphabet.
In practice, this is non-trivial to implement, requires significant preprocessing time and working memory and the overhead caused by backward searches is not satisfactory.

Kempa~and~Kosolobov~\cite{DBLP:conf/esa/KempaK17} showed that the LZ-End parsing can be computed in time $\Oh{n}$ using a sophisticated approach that we conjecture to be impractical.
However, they also present an intermediate algorithm that requires $\Oh{n\lg\lg n}$ and a much smaller set of tools.
They have furthermore implemented this algorithm in the \emph{lz-end-toolkit}\footnote{\url{https://github.com/dominikkempa/lz-end-toolkit}} to accompany the article~\cite{DBLP:conf/dcc/KempaK17}, where it is used as a block-local parser within a larger-scale streaming algorithm.
The log-logarithmic factor in the running time comes from predecessor and successor queries, which are indeed the main bottleneck.
However, with very efficient data structures being available for these kinds of queries (see, e.g., \cite{DBLP:conf/esa/Dinklage0HKK20}), the algorithm is of high practical interest.

In this work, we apply practical improvements to the $\Oh{n\lg\lg n}$-time algorithm that at the same time simplify it.
First, we significantly reduce the number of queries and modifications on the predecessor/successor data structure by doing them lazily as needed.
Second, we make the data structure associative (i.e., an ordered dictionary), which allows us to discard the suffix array after initialization.
This eliminates a level of indirection and reduces the required amount of working memory.
Finally, we adopt some undocumented tricks from the aforementioned implementation of~\cite{DBLP:conf/dcc/KempaK17}.
The resulting algorithm is simple and easy to implement.
We give a complete listing and an in-depth description in section~\ref{sect:algorithm} including examples.
In section~\ref{sect:impl}, we demonstrate the practical relevancy of the algorithm.

\section{Preliminaries}
\label{sect:prelims}

We argue about time and space requirements using the \emph{word RAM} model \cite{DBLP:conf/stacs/Hagerup98}, where the memory consists of contiguous words of length $\Theta(\lg n)$ bits each (by default, we state logarithms as base-2).
We can access and perform arithmetic operations on a constant number of words in constant time.

For a string $S \in \Sigma^n$ of length $n$ and some $i \in [0, n-1]$, we denote by $S[i]$ the $i$-th character of $S$.
We deliberately start indexing at zero to remain practical.
For a $j \in [i,n-1]$, we denote by $S[i~\twodots~j]$ the substring of $S$ beginning at position $i$ and ending at position $j$, both included.
We denote the concatenation of two strings $u, v \in \Sigma^*$ by the juxtaposition $uv$.

\subsection{The LZ-End Parsing}
\label{sect:lz-end}

Let $S \in \Sigma^n$ be a string of length $n$. 
The LZ-End parsing is a concatenation of $z$ phrases $f_1, \dots, f_z$ such that $f_1 \cdots f_z = S$.
For $p \in [1,z]$, we define $S_{\overrightarrow{p}} := f_1 \cdots f_p = S[0 \twodots |f_1|+\cdots+|f_p|-1]$.
Each phrase $f_i$ in the LZ-End parsing (for $i \in [1,z]$) is a string $s_p \alpha$ such that
(1)~$s_p$ is the suffix of some $S_{\overrightarrow{p}}$ with $p < i$ and
(2)~there is no $p' < i$ such that $s_p \alpha$ is a suffix of $S_{\overrightarrow{p'}}$.
We define $f_0 := \varepsilon$ such that $p = 0$ and $s_p = \varepsilon$ are a valid default case for condition (1).
Intuitively, an LZ-End phrase extends the longest possible previous string that \emph{ends} at a phrase boundary, hence the name.
In this sense, it is also a greedy parsing.

We represent the phrase $f_i$ as the triple $(p, \ell, \alpha)$,
where $p$ is the index of the \emph{source phrase},
$\alpha$ is the \emph{extension}
and $\ell := |f_i|$ is the \emph{length} of the phrase (including $\alpha$).

\begin{example}
\label{ex:lz-end}
We parse the string $S = \texttt{a}\,\texttt{b}\,\texttt{aa}\,\texttt{baa\$}$ ($n=8$) phrase by phrase.
\begin{enumerate}
\item The first phrase is trivially $f_1 = (0,1,\texttt{a})$ because there is no suitable source phrase.
\item Similarly, the second phrase $f_2 = (0,1,\texttt{b})$.
\item The next phrase is $f_3 = (1, 2, \texttt{a})$, because the longest previous string that ends at a phrase boundary is \texttt{a} (length-1 suffix of $S_{\overrightarrow{1}} = \texttt{a}$) and we extend it by the next character \texttt{a}.
\item The final phrase is $f_4 = (3, 4, \texttt{\$})$, because we can copy the length-3 string \texttt{baa} ending at the boundary of $f_3$ (length-3 suffix of $S_{\overrightarrow{3}} = \texttt{abaa}$) and append the final character \texttt{\$}.
\end{enumerate}
The size of the LZ-End parsing of $S$ is therefore $z=5$.
\end{example}

\subsection{Index Data Structures}
\label{sect:index}

The \emph{suffix array} $A$ conceptually contains the suffixes of $S$ in lexicographic order~\cite{DBLP:journals/siamcomp/ManberM93}.
We only actually store the starting positions of said suffixes in such that for $i \in [0,n-1]$, the suffix $S[A[i]~\twodots~n-1]$ is the lexicographically $(i+1)$-smallest suffix of $S$.
It is well known that the suffix array can be computed in time and space linear in the input size~\cite{DBLP:journals/tc/NongZC11}.
Unlike \cite{DBLP:conf/esa/KempaK17}, we will not need the suffix array directly to compute the LZ-End parsing, but only during preprocessing to compute its inverse and the LCP array.

The \emph{inverse suffix array} $A^{-1}$ is the inverse permutation of $A$.
For $i \in [0, n-1]$, $A^{-1}[i]$ stores the lexicographic rank of the suffix $S[i~\twodots~n-1]$, i.e., its position in $A$.
It thus holds that $A^{-1}[A[i]] = i$, making it trivial to compute in time $\Oh{n}$ if $A$ has already been computed.

For two integers $i, j \in [0,n-1]$ with $i \leq j$, we define the \emph{longest common extension}
\begin{align*}
    \lce(i,j) := \max\{ \ell ~\vert~ S[i~\twodots~i+\ell] = S[j~\twodots~j+\ell] ~\land~ S[i+\ell+1] \neq S[j+\ell+1] \}.
\end{align*}

In the \emph{LCP array} $H$, we store the length of the longest common prefix between two neighboring suffixes in $A$.
To this end, it is $H[i] := \lce(A[i], A[i-1])$ for $i \in [1, n-1]$.
For a complete definition, we set $H[0] := 0$.
From the suffix array, the LCP array can be computed time $\Oh{n}$ using the popular and simple algorithm by Kasai et al.~\cite{DBLP:conf/cpm/KasaiLAAP01}.

It is well known that given $H$ and $A^{-1}$, we can compute $\lce(i, j)$ using a \emph{range minimum query} (rmq) over $H$.
Without loss of generality, let $A^{-1}[i] < A^{-1}[j]$ (i.e., the suffix of $S$ starting at position $i$ is lexicographically smaller than the one starting at $j$).
Then,
\begin{align*}
\lce(i, j) &= \rmq(H, A^{-1}[i]+1, ~A^{-1}[j]) \\
&:= \text{argmin}\{ H[x] ~\vert~ x \in [A^{-1}[i]+1, ~A^{-1}[j]] \}.
\end{align*}

In time $\Oh{n}$, we can compute a linear-space data structure that answers range minimum queries in constant time~\cite{DBLP:journals/siamcomp/FischerH11}.
Clearly, storing $A$, $A^{-1}$ and $H$ plainly requires $\Oh{n}$ space.

\subsection{Dynamic Associative Predecessor and Successor Data Structures}
\label{sect:pred}

Let $X \subseteq U$ be a set of \emph{keys} drawn from some universe $U = [0, u-1] \subseteq \setN$ of size $u$.
For some $y \in U$, the \emph{predecessor} $\qpred(y) := \max\{ x \in X ~\vert~ x \leq y \}$ is the largest integer contained in $X$ that is less than or equal to $y$.
Symmetrically, the \emph{successor} $\qsucc(y) := \min\{ x \in X ~\vert~ x \geq y \}$ is the smallest integer contained in $X$ that is greater than or equal to $y$.
The problem of finding predecessors or successors is well studied both in theory and practice.
We consider the \emph{dynamic} case where we want to answer both queries efficiently while $X$ is subject to insertions and deletions, which we also need to do efficiently.
Navarro and Rojas-Ledesma~\cite{NavarroR20} give a thorough survey whereas a recent practical study is done by \cite{DBLP:conf/wea/Dinklage0H21}.

Balanced search trees are a simple and widely available data structure for the problem.
Their space usage is $\Oh{|X|}$ and insertions, deletions, predecessor and successor queries can be performed or answered in time $\Oh{\lg |X|}$, respectively.
In the light of computing the LZ-End parsing, we will have that $|X| = \Th{z}$, and thus any search tree will satisfy a space bound of $\Oh{z}$.
We will also have that $u = n$, because the universe of keys in $X$ consists of pointers into the input string $S$.
This motivates the use of length-reducing data structures such as the van Emde Boas tree~\cite{DBLP:conf/focs/Boas75} or the y-fast trie~\cite{DBLP:journals/ipl/Willard83} as done in \cite{DBLP:conf/esa/KempaK17} and \cite{DBLP:conf/dcc/KreftN10}, respectively,
offering $\Oh{\lg\lg n}$-time operations and queries while consuming space $\Oh{n}$.

We will require the data structure to be \emph{associative}, redefining $X \subseteq U \times D$ for some universe $D$ of satellite data.
Predecessor and successor then pertain to the \emph{keys} from $U$ and the associated data from $D$ is returned along with the result.
The aforementioned data structures can all be extended to store satellite data, essentially making them ordered dictionaries.
In our scenario, we will have that $D = [1,z]$, and thus their space bound is retained.
For a pair $(x,d) \in X$ we also write $x \mapsto d$.

\section{Computing the LZ-End Parsing}
\label{sect:algorithm}

Given an input string $S \in \Sigma^n$, we want to compute the LZ-End parsing $f_1 \cdots f_z$ of $S$.
Let us first recapture the following observation by Kempa and Kosolobov~\cite[Lemma~3]{DBLP:conf/esa/KempaK17}.

\begin{lemma}
\label{lemma:kk17}
If $f_1 \cdots f_z$ is the LZ-End parsing of a string $S \in \Sigma^*$, then, for any character $\alpha \in \Sigma$,
the last phrase in the LZ-End parsing of $S\alpha$ is (1)~$f_{z-1} f_z \alpha$ or (2)~$f_z \alpha$ or (3) $\alpha$.
\end{lemma}

Essentially, this allows us to compute the LZ-End parsing in a left-to-right scan of $S$ where in each step,
we only have to consider to either
(1)~\emph{merge} the two most recent phrases,
(2)~\emph{extend} the most recent phrase or
(3)~\emph{begin} a new phrase consisting of a single character.

Suppose that we already parsed the prefix $S[0 \twodots i-1]$ for some position $i > 0$ and let $f_1 \cdots f_z$ be the LZ-End parsing thus far.
In the next step, we compute the LZ-End parsing of the prefix $S[0 \dots i]$.
We look for a suffix $s_p$ of $S[0 ~\twodots~ i-1]$ that ends at the boundary of some source phrase $p < z$ (i.e., it is also a suffix of $S_{\overrightarrow{p}}$) and has length $|s_p| \geq |f_z|$.
Then, we greedily decide which of the three aforementioned cases applies.
If $|s_p| \geq |f_{z-1}|+|f_z|$, it means that the new phrase covers at least the two most recent phrases and we can merge them to $(p, |f_{z-1}|+|f_z|+1, S[i])$.
Otherwise, if $|s_p| \geq |f_z|$, we can extend the most recent phrase to $(p, |f_z|+1, S[i])$.
If neither applies, we begin a new phrase $(0,1,S[i])$.

\begin{figure}
\begin{center}
\begin{tikzpicture}[
    every node/.style={
        anchor=south,
    },
    char/.style={
        font=\ttfamily,
    },
    gray/.style={
        color=gray,
    },
    cursor/.style={
        font=\small,
        minimum width=5em,
    },
    index/.style={
        font=\small,
    },
    action/.style={
        font=\small,
        minimum width=15em,
    },
]

\matrix[
    matrix of nodes,
    column sep=1ex,
]{
    \node[cursor](head-first){Step}; &
    \node[index]{0}; &
    \node[index]{1}; &
    \node[index]{2}; &
    \node[index]{3}; &
    \node[index]{4}; &
    \node[index]{5}; &
    \node[index]{6}; &
    \node[index]{7}; &
    \node[action](head-last){Action}; \\
    \node[cursor]{$i=0$}; &
    \node[char](i0-cur){a}; &
    \node[char,gray]{b}; &
    \node[char,gray]{a}; &
    \node[char,gray]{a}; &
    \node[char,gray]{b}; &
    \node[char,gray]{a}; &
    \node[char,gray]{a}; &
    \node[char,gray]{\$}; &
    \node[action]{begin $f_1 := (0,1,\texttt{a})$}; \\
    \node[cursor]{$i=1$}; &
    \node[char](i1-f1){a}; &
    \node[char](i1-cur){b}; &
    \node[char,gray]{a}; &
    \node[char,gray]{a}; &
    \node[char,gray]{b}; &
    \node[char,gray]{a}; &
    \node[char,gray]{a}; &
    \node[char,gray]{\$}; &
    \node[action]{begin $f_2 := (0,1,\texttt{b})$}; \\
    \node[cursor]{$i=2$}; &
    \node[char](i2-f1){a}; &
    \node[char](i2-f2){b}; &
    \node[char](i2-cur){a}; &
    \node[char,gray]{a}; &
    \node[char,gray]{b}; &
    \node[char,gray]{a}; &
    \node[char,gray]{a}; &
    \node[char,gray]{\$}; &
    \node[action]{begin $f_3 := (0,1,\texttt{a})$}; \\
    \node[cursor]{$i=3$}; &
    \node[char](i3-f1){a}; &
    \node[char](i3-f2){b}; &
    \node[char](i3-f3){a}; &
    \node[char](i3-cur){a}; &
    \node[char,gray]{b}; &
    \node[char,gray]{a}; &
    \node[char,gray]{a}; &
    \node[char,gray]{\$}; &
    \node[action]{extend $f_3 := (1,2,\texttt{a})$}; \\
    \node[cursor]{$i=4$}; &
    \node[char](i4-f1){a}; &
    \node[char](i4-f2){b}; &
    \node[char](i4-f3){a}; &
    \node[char](i4-f3-end){a}; &
    \node[char](i4-cur){b}; &
    \node[char,gray]{a}; &
    \node[char,gray]{a}; &
    \node[char,gray]{\$}; &
    \node[action]{begin $f_4 := (0,1,\texttt{b})$}; \\
    \node[cursor]{$i=5$}; &
    \node[char](i5-f1){a}; &
    \node[char](i5-f2){b}; &
    \node[char](i5-f3){a}; &
    \node[char](i5-f3-end){a}; &
    \node[char](i5-f4){b}; &
    \node[char](i5-cur){a}; &
    \node[char,gray]{a}; &
    \node[char,gray]{\$}; &
    \node[action]{extend $f_4 := (2,2,\texttt{a})$}; \\
    \node[cursor]{$i=6$}; &
    \node[char](i6-f1){a}; &
    \node[char](i6-f2){b}; &
    \node[char](i6-f3){a}; &
    \node[char](i6-f3-end){a}; &
    \node[char](i6-f4){b}; &
    \node[char](i6-f4-end){a}; &
    \node[char](i6-cur){a}; &
    \node[char,gray]{\$}; &
    \node[action]{begin $f_5 := (0,1,\texttt{a})$}; \\
    \node[cursor]{$i=7$}; &
    \node[char](i7-f1){a}; &
    \node[char](i7-f2){b}; &
    \node[char](i7-f3){a}; &
    \node[char](i7-f3-end){a}; &
    \node[char](i7-f4){b}; &
    \node[char]{a}; &
    \node[char]{a}; &
    \node[char](i7-cur){\$}; &
    \node[action]{merge $f_4$ and $f_5$ to $f_4:= (3,4,\texttt{\$})$}; \\
};

\draw[black] (head-first.south west) to (head-last.south east);
\draw[black,very thick] ([xshift=-0.5ex]i0-cur.north west) to ([xshift=-0.5ex]i0-cur.south west);
\draw[black,very thick] ([xshift=-0.5ex]i1-cur.north west) to ([xshift=-0.5ex]i1-cur.south west);
\draw[black,very thick] ([xshift=-0.5ex]i2-cur.north west) to ([xshift=-0.5ex]i2-cur.south west);
\draw[black,very thick] ([xshift=-0.5ex]i3-cur.north west) to ([xshift=-0.5ex]i3-cur.south west);
\draw[black,very thick] ([xshift=-0.5ex]i4-cur.north west) to ([xshift=-0.5ex]i4-cur.south west);
\draw[black,very thick] ([xshift=-0.5ex]i5-cur.north west) to ([xshift=-0.5ex]i5-cur.south west);
\draw[black,very thick] ([xshift=-0.5ex]i6-cur.north west) to ([xshift=-0.5ex]i6-cur.south west);
\draw[black,very thick] ([xshift=-0.5ex]i7-cur.north west) to ([xshift=-0.5ex]i7-cur.south west);

\draw[black] (i0-cur.south west) to (i0-cur.south east);

\draw[black] (i1-f1.south west) to (i1-f1.south east);
\draw[black] (i1-cur.south west) to (i1-cur.south east);

\draw[black] (i2-f1.south west) to (i2-f1.south east);
\draw[black] (i2-f2.south west) to (i2-f2.south east);
\draw[black] (i2-cur.south west) to (i2-cur.south east);

\draw[black] (i3-f1.south west) to (i3-f1.south east);
\draw[black] (i3-f2.south west) to (i3-f2.south east);
\draw[black] (i3-f3.south west) to (i3-cur.south east);

\draw[black] (i4-f1.south west) to (i4-f1.south east);
\draw[black] (i4-f2.south west) to (i4-f2.south east);
\draw[black] (i4-f3.south west) to (i4-f3-end.south east);
\draw[black] (i4-cur.south west) to (i4-cur.south east);

\draw[black] (i5-f1.south west) to (i5-f1.south east);
\draw[black] (i5-f2.south west) to (i5-f2.south east);
\draw[black] (i5-f3.south west) to (i5-f3-end.south east);
\draw[black] (i5-f4.south west) to (i5-cur.south east);

\draw[black] (i6-f1.south west) to (i6-f1.south east);
\draw[black] (i6-f2.south west) to (i6-f2.south east);
\draw[black] (i6-f3.south west) to (i6-f3-end.south east);
\draw[black] (i6-f4.south west) to (i6-f4-end.south east);
\draw[black] (i6-cur.south west) to (i6-cur.south east);

\draw[black] (i7-f1.south west) to (i7-f1.south east);
\draw[black] (i7-f2.south west) to (i7-f2.south east);
\draw[black] (i7-f3.south west) to (i7-f3-end.south east);
\draw[black] (i7-f4.south west) to (i7-cur.south east);

\end{tikzpicture}
\end{center}
\caption{
    Character-wise LZ-End parsing of the string $S = \texttt{a}\,\texttt{b}\,\texttt{aa}\,\texttt{baa\$}$,
    applying the case distinction of Lemma~\ref{lemma:kk17} in every step.
    Refer to Example~\ref{ex:kk} for a description of the individual steps.
}
\label{fig:ex-kk}
\end{figure}

\begin{example}
\label{ex:kk}
We parse the string $S = \texttt{a}\,\texttt{b}\,\texttt{aa}\,\texttt{baa\$}$ ($n=8$) from Example~\ref{ex:lz-end} character by character using the aforementioned case distinction in every step.
Figure~\ref{fig:ex-kk} shows the state in the individual steps.
\begin{enumerate}
\item In the first step, we trivially set $f_1 := (0,1,\texttt{a})$ as there cannot be any source phrase.
\item We start a new phrase $f_2 := (0,1,\texttt{b})$ because no suffix of the already parsed string \texttt{a} has an occurrence prior to $f_1$.
\item The next phrase is initially a new phrase $f_3 := (0,1,\texttt{a})$, because no suffix of the already parsed string \texttt{ab} has an occurrence prior to $f_2$.
\item Now, we can extend the phrase to $f_3 := (1,2,\texttt{a})$ because the length-1 suffix \texttt{a} of the already parsed string \texttt{aba} has a previous occurrence ending at the boundary of phrase $f_1$.
\item We cannot extend $f_3$ any further, because no suffix of the already parsed string \texttt{abaa} has an occurrence prior to $f_3$.
      Therefore, we begin a new phrase $f_4 := (0,1,\texttt{b})$.
\item We extend $f_4 := (2,2,\texttt{a})$ because the length-1 suffix \texttt{b} of the already parsed string \texttt{abaab} has a previous occurrence ending at the boundary of phrase $f_2$.
\item Now, we cannot extend $f_4$ any further.
      Even though the suffix \texttt{aba} of the already parsed string \texttt{abaaba} has a previous occurrence prior to $f_4$, it does not coincide with a phrase boundary.
      Therefore, we have to begin the new phrase $f_5 := (0,1,\texttt{a})$.
\item Now, however, we can merge $f_4$ and $f_5$ to the final phrase $f_4 = (3,4,\$)$, because the length-3 suffix \texttt{baa} of the already parsed string \texttt{abaabaa} has an occurrence ending at the boundary of phrase $f_3$.
\end{enumerate}
The resulting parsing is the same as in Example~\ref{ex:lz-end}.
\end{example}

\subsection{Finding Source Phrases}

The core problem when computing the parsing is finding a suitable source phrase to copy from.
We use an index consisting of the inverse suffix array $A^{-1}$ and the LCP array $H$ (with constant-time rmq support) over the reverse input $\overleftarrow{S}$.
Additionally, we maintain a dynamic predecessor/successor data structure $M$ in which we associate the ending positions of already computed LZ-End phrases with the corresponding phrase number.
The positions contained in $M$ are in the lex-space of the suffixes of $\overleftarrow{S}$: if LZ-End phrase $f_x$ ends at position $i$ in $S$, then we insert $A^{-1}[n-i-1] \mapsto x$ into $M$.
Source phrase candidates can then be found using predecessor and successor queries on $M$, while the possible number of characters that can be copied from these candidates can be computed via range-minimum queries on $H$.
Whichever candidate allows for a sufficiently long copy corresponds to a string $s_p$ as described earlier.
Thanks to the associative nature of $M$, unlike \cite{DBLP:conf/esa/KempaK17}, we never consider positions in $S$, completely eliminating the need to access the suffix array $A$.

The index is computed for the \emph{reverse} input $\overleftarrow{S}$, but we do process $S$ from left to right,
Like in the implementaion of \cite{DBLP:conf/dcc/KempaK17}, we use a variant $A'$ of the inverse suffix array such that $A'[A[n-i-1]] := i$, thus, $A'[i] = A^{-1}[n-i-1]$.
This saves us the frequent arithmetics needed to transform positions between $S$ and $\overleftarrow{S}$ and makes the algorithm more comprehensible.

\subsection{The Algorithm}

\begin{algorithm}[p!]
\DontPrintSemicolon
\SetKwFunction{FParse}{\textsc{Parse}}
\SetKwFunction{FLeftPhrase}{\textsc{LexSmallerPhrase}}
\SetKwFunction{FRightPhrase}{\textsc{LexGreaterPhrase}}
\SetKwFunction{FFindCandidate}{\textsc{FindCopySource}}
\begin{small}
\Fn{\FParse{$S$}}{
    \tcc{build index and initialize}
    $A \gets $ suffix array of $\overleftarrow{S}$,
    $H \gets $ LCP array of $\overleftarrow{S}$\ (with rmq support)\;
    $A' \gets $ array of length $n$\;
    \lFor{$i \gets 0$ \KwTo $n-1$}{
        $A'[n-A[i]-1] = i$
    }
    discard $A$ and $\overleftarrow{S}$\;
    $f_0 \gets (0,~0,~\varepsilon)$,
    $f_1 \gets (0,~1,~S[0])$,
    $z \gets 1$,
    $M \gets \emptyset$\;
    \For{$i \gets 1$ \KwTo $n-1$}{\label{algo:parse:main-loop}
        $i' \gets A'[i-1]$ \tcp{suffix array neighbourhood of $i-1$ in $\overleftarrow{S}$}
        $p_1 \gets \bot$,
        $p_2 \gets \bot$\;
        \tcc{find candidates}
        \FFindCandidate{\FLeftPhrase}\;\label{algo:parse:left}
        \If{$p_1 = \bot$ \Or $p_2 = \bot$}{
            \FFindCandidate{\FRightPhrase}\;\label{algo:parse:right}
        }
        \tcc{case distinction according to Lemma~\ref{lemma:kk17}}
        \If{$p_2 \neq \bot$}{
            \tcc{merge phrases $f_{z-1}$ and $f_z$}
            $M \gets M \setminus \{ (A'[i-|f_z|-1], \ast) \}$ \tcp{unmark phrase $f_{z-1}$}\label{algo:parse:unmark}
            $f_{z-1} \gets (p_2, ~|f_z| + |f_{z-1}| + 1, ~S[i])$\;
            $z \gets z-1$\;
        }\ElseIf{$p_1 \neq \bot$}{
            \tcc{extend phrase $f_z$}
            $f_{z} \gets (p_1, ~|f_z| + 1, ~S[i])$\;
        }\Else{\label{algo:parse:new}
            \tcc{begin new phrase}
            $M \gets M \cup \{ i', z \}$ \tcp{lazily mark phrase $f_z$}
            $f_{z+1} \gets (0, ~1, ~S[i])$\;
            $z \gets z+1$\;
        }
    }
    \Return $(f_1, \dots, f_z)$\;
}
\;
\Fn{\FFindCandidate{$f$}}{
    $(j', p, \ell) \gets f(i')$\;
    \If{$\ell \geq |f_z|$}{
        $p_1 \gets p$\;\label{algo:parse:extend}
        \If{$i > |f_z|$}{
            \lIf{$p = z-1$}{
                $(j', p, \ell_L) \gets f(j')$
            }\label{algo:parse:left-phrase2}
            \lIf{$\ell \geq |f_z| + |f_{z-1}|$}{
                $p_2 \gets p$\label{algo:parse:merge}
            }
        }
    }
}
\;
\Fn{\FLeftPhrase{$i'$}}{
    \If{$M$ contains the predecessor $j' \leq i'-1$ with $j' \mapsto p$}{
        \Return $(j',p, H[\rmq(j'+1, i')])$\;
    }\lElse{
        \Return $(0,0,0)$
    }
}
\;
\Fn{\FRightPhrase{$i'$}}{
    \If{$M$ contains the successor $j' \geq i'+1$ with $j' \mapsto p$}{
        \Return $(j',p, H[\rmq(i'+1, j')])$\;
    }\lElse{
        \Return $(0,0,0)$
    }
}
\end{small}
\vspace{1ex}
\caption{
    Algorithm to compute the LZ-End parsing for a string $S \in \Sigma^n$ (function \textsc{Parse}).
}
\label{algo:parse}
\end{algorithm}

Algorithm~\ref{algo:parse} lists our algorithm to compute the LZ-End parsing for an input string $S$.
We begin by building our index over the reversed input $\overleftarrow{S}$.
The first LZ-End phrase is always $f_1 = (0, 1, S[0])$, and so we initialize the number $z$ of phrases as 1.
We do \emph{not} mark the end position of $f_1$ in $M$ yet.
Rather, $M$ will be updated lazily: whenever a new phrase begins, we insert the previous phrase $f_z$ into $M$, and we only ever remove a phrase ($f_{z-1}$) from $M$ if two phrases are merged.
This is unlike the algorithm described in~\cite{DBLP:conf/esa/KempaK17}, as we will discuss later.
Thus, we initialize $M$ to be empty.
The empty phrase $f_0$ is initialized only for technical reasons (such that $f_{z-1}$ is always well-defined).

We proceed to parse $S$ from left to right in the main loop beginning in line~\ref{algo:parse:main-loop}.
As described earlier, at position $i$, we look for a suffix of $S[0~\twodots~i-1]$ that has an occurrence ending at the boundary of some phrase $f_p$ prior to the most recent phrase $f_z$.
If the suffix is sufficiently long, we can use $f_p$ as a source to copy from to either extend $f_z$ or even merge $f_z$ and $f_{z-1}$.
The search is done in the suffix array neighbourhood of $i' = A'[i-1]$ in $\overleftarrow{S}$.

The main tools are provided by the functions \textsc{LexSmallerPhrase} and \textsc{LexGreaterPhrase}, which report the lexicographically closest (less or greater, respectively) suffix of $\overleftarrow{S}$ that is also the boundary of a previous phrase.
In \textsc{LexSmallerPhrase}, this is done using a predecessor search in $M$ starting from $i'-1$ (explicitly excluding position $i'$, which is the boundary of the most recent phrase $f_z$).
If a predecessor is found at position $j'$ marking the boundary of phrase $f_p$, we compute the maximum possible copy length $\ell$ using a constant-time range minimum query on the LCP array $H$ and report the triple $(j', p, \ell)$.
The function \textsc{LexGreaterPhrase} symmetrically does a successor search in $M$.

\begin{figure}[t]
\begin{center}
\begin{tikzpicture}[
    every node/.style={
        font=\small,
    },
    box/.style={
        draw,
        shape=rectangle,
        minimum height=1.5em,
        minimum width=1.25em
    },
    dots/.style={
        box,
        dotted,
        minimum width=5em,
    },
    virt/.style={
        minimum height=1.5em,
    }
]
\node(llldots){$\cdots$};
\node[left=0 of llldots]{$A$};
\node[box,right=0 of llldots](pl2){~};
\node[dots,right=0 of pl2](lldots){$\cdots$};
\node[box,right=0 of lldots](pl1){~};
\node[below=1em of pl1](pl1-label){$p_L$};
\node[dots,right=0 of pl1](ldots){$\cdots$};
\node[box,right=0 of ldots](z){~};
\node[below=1em of z](z-label){$z$};
\node[dots,right=0 of z](rdots){$\cdots$};
\node[box,right=0 of rdots](pr1){~};
\node[below=1em of pr1](pr1-label){$p_R$};
\node[dots,right=0 of pr1](rrdots){$\cdots$};
\node[box,right=0 of rrdots](pr2){~};
\node[right=0 of pr2]{$\cdots$};

\draw[solarized-blue,fill=solarized-blue] (pl2) circle (.25em);
\draw[solarized-blue,fill=solarized-blue] (pl1) circle (.25em);
\draw[solarized-blue] (pl1-label) -- (pl1.center);
\draw[solarized-blue] (z) circle (.25em);
\draw[solarized-blue,dashed] (z-label) -- (z.center);
\draw[solarized-blue,fill=solarized-blue] (pr1) circle (.25em);
\draw[solarized-blue] (pr1-label) -- (pr1.center);
\draw[solarized-blue,fill=solarized-blue] (pr2) circle (.25em);

\node[above=2em of z](i){$i'$};
\draw[solarized-red,-Latex] (i) -- (z);

\node[above=2em of pl1](jl1){$j'_L$};
\draw[solarized-yellow,-Latex] (jl1) -- (pl1);
\draw[solarized-yellow,-Latex] (i.north west) to[out=135,in=45,looseness=0.8] node[yshift=0.5em,xshift=-0.5em]{\scriptsize\textsc{LexSmallerPhrase}} (jl1.north);

\node[virt,above=1.5em of pl2](jl2){~};
\draw[solarized-green,dotted,-Latex] (jl1.north east) to[out=135,in=45,looseness=0.8] (jl2.north);
\draw[solarized-green,dotted,-Latex] (jl2.north) -- (pl2);

\node[above=1.5em of pr1](jr1){$j'_R$};
\draw[solarized-yellow,-Latex] (jr1) -- (pr1);
\draw[solarized-yellow,-Latex] (i.north east) to[out=45,in=135,looseness=0.8] node[yshift=0.5em,xshift=0.5em]{\scriptsize\textsc{LexGreaterPhrase}} (jr1.north);

\node[virt,above=1.5em of pr2](jr2){~};
\draw[solarized-green,dotted,-Latex] (jr1.north east) to[out=45,in=135,looseness=0.8] (jr2.north);
\draw[solarized-green,dotted,-Latex] (jr2.north) -- (pr2);
\end{tikzpicture}
\end{center}
\caption{
    Search for copy source phrase candidates in suffix array space (\textsc{FindCopySource}).
    Positions that mark the ending locations in $M$ of already computed LZ-End phrases are indicated by the circles.
    Note that while $i'$ is the ending location of the most recent phrase $f_z$, it has not yet been entered into $M$.
    We do a predecessor query starting from $i'-1$ (\textsc{LexSmallerPhrase}) or a successor query starting from $i'+1$ (\textsc{LexGreaterPhrase}), giving us the locations $j'_L$ or $j'_R$ that mark the candidates $p_L$ and $p_R$, respectively.
    In case $p_L = z-1$ or $p_R = z-1$, we do another predecessor or successor query starting from $j'_L-1$ or $j'_R+1$, respectively, to find a candidate for merging.
    Using range minimum queries in the LCP array, we can find the longest common extension between the suffix of $\overleftarrow{S}$ starting at $A[i']$ and those starting at $A[j'_L]$ or $A[j'_R]$, respectively, which is the number of characters that can be copied from the corresponding source phrase.
}
\label{fig:candidates}
\end{figure}

For merging, since we merge phrases $f_z$ and $f_{z-1}$, it is important that $p \neq z-1$ (we cannot copy from a phrase that we merge away).
Therefore, we distinguish between the source phrase $p_1$ for a possible extension of $f_z$ and the source phrase $p_2 \neq z-1$ that can potentially be used to merge $f_z$ and $f_{z-1}$.
The function \textsc{FindCopySource}, given a directional search function $f$ -- either \textsc{LexSmallerPhrase} or \textsc{RightSmallerPhrase} -- takes care of this distinction.
We use $f$ to find the triple $(j', p, \ell)$ as described earlier.
In case $\ell \geq |f_z|$, the phrase $f_p$ is a suitable copy source for extending $f_z$ and we set $p_1 := p$ (line~\ref{algo:parse:extend}).
If $p = z-1$, we use $f$ again, this time starting off (and excluding) position $j'$ that marks the boundary of phrase $f_{z-1}$.
If then, $\ell \geq |f_z|+|f_{z-1}|$, the phrase $f_p$ is a suitable copy source for merging $f_z$ and $f_{z-1}$ and we set $p_2 := p$ (line~\ref{algo:parse:merge}).
Figure~\ref{fig:candidates} visualizes the described search.

In our main loop, we first call \textsc{FindCopySource} with the directional search function \textsc{LexSmallerPhrase} to find lexicographically smaller extension and/or merging candidates (line~\ref{algo:parse:left}).
In case none were found, we also look for lexicographically larger candidates (line~\ref{algo:parse:right}).
We then proceed with the case distinction according to Lemma~\ref{lemma:kk17}.
If a suitable merge candidate was found ($p_2 \neq \bot$), we merge phrases $f_{z}$ and $f_{z-1}$ into a phrase of length $|f_z| + |f_{z-1}| + 1$ with source phrase $p_2$.
In this case, we also need to permanently delete the ending position of phrase $f_{z-1}$ from $M$ and decrease the number $z$ of current phrases.
Otherwise, if an extension candidate was found ($p_1 \neq \bot$), we extend the most recent phrase $f_z$ to length $|f_z|+1$ with source phrase $p_1$.
If neither applies, we begin a new phrase of length $1$ consisting of only the character $S[i]$.
Only then we lazily mark the ending position of the previous phrase $f_z$ in $M$ and subsequently increment $z$.
After function \textsc{Parse} has finished, the phrases $f_1, \dots, f_z$ contain the LZ-End parsing of $S$ and are returned.

\begin{figure}
\begin{center}
\begin{tikzpicture}[
    every node/.style={
        font=\small,
    },
    index/.style={
        font=\scriptsize,
    },
    char/.style={
        font=\ttfamily,
    },
    int/.style={
    },
]
\matrix[
    matrix of nodes,
    column sep=1em,
]{
    \node[index](head-first){}; &
    \node[index]{0}; &
    \node[index]{1}; &
    \node[index]{2}; &
    \node[index]{3}; &
    \node[index]{4}; &
    \node[index]{5}; &
    \node[index]{6}; &
    \node[index]{7}; \\
    \node(head1){$\overleftarrow{S}$}; &
    \node[char]{\$}; &
    \node[char]{a}; &
    \node[char]{a}; &
    \node[char]{b}; &
    \node[char]{a}; &
    \node[char]{a}; &
    \node[char]{b}; &
    \node(head2)[char]{a}; \\
    $A$ &
    \node[int]{0}; &
    \node[int]{7}; &
    \node[int]{4}; &
    \node[int]{1}; &
    \node[int]{5}; &
    \node[int]{2}; &
    \node[int]{6}; &
    \node[int]{3}; \\
    $H$ &
    \node[int]{0}; &
    \node[int]{0}; &
    \node[int]{1}; &
    \node[int]{4}; &
    \node[int]{1}; &
    \node[int]{3}; &
    \node[int]{0}; &
    \node[int]{2}; \\
    $M$ &
    ~ &
    \node(f1){$f_1$}; &
    \node(f3){$f_3$}; &
    \node[color=gray](f5)[draw=black,circle,inner sep=0.5ex]{$f_5$}; &
    ~ &
    \node(f4){$f_4$}; &
    \node(f2){$f_2$}; &
    ~ \\
};

\draw (head1.south west) to (head2.south east);
\draw[-Latex,thick] (f5.south) to[out=-135,in=-45] (f3.south);
\draw[red,dashed] (f4.north west) to (f4.south east);
\draw[red,dashed] (f4.north east) to (f4.south west);
\end{tikzpicture}
\end{center}
\caption{
    The relevant data structures of Algorithm~\ref{algo:parse} right before merging $f_5$ and $f_4$ in the final step of parsing $S = \texttt{a}\,\texttt{b}\,\texttt{aa}\,\texttt{ba}\,\texttt{abaa\$}$.
    $A$ and $H$ are the suffix and LCP array of $S$, respectively, and we associate a position $M[i]$ with a phrase iff that phrase ends at position $A[i]$ in $\overleftarrow{S}$.
    Note that $\overleftarrow{S}$ and $A$ are not actually stored but only shown for reference.
    The phrase $f_3$ is found as a suitable source phrase, allowing to copy up to $\rmq_H(3,3)=4$ characters.
    As a result of the merge, the former boundary of phrase $f_4$ is removed from $M$.
    Refer to Example~\ref{ex:merge} for an elaboration.
}
\label{fig:merge}
\end{figure}

\begin{example}
\label{ex:merge}
Figure~\ref{fig:merge} shows the scenario of Algorithm~\ref{algo:parse} right before the merging phrases $f_5$ and $f_4$ in the final step of parsing $S = \texttt{a}\,\texttt{b}\,\texttt{aa}\,\texttt{ba}\,\texttt{abaa\$}$ as in Example~\ref{ex:kk}.
We already parsed the first seven characters of $S$ and thus start our search in $M$ for source phrases from position $A'[i-1]=A'[6]=3$ (because $S[6]$ corresponds to $\overleftarrow{S}[1]$ and it is $A[3] = 1$).
The result of \textsc{FindCopySource} with directional search function \text{LexSmallerPhrase} (line~\ref{algo:parse:left}) is the triple $(j', p, \ell) = (2, 3, 4)$, meaning that we can copy up to 4 characters from the phrase boundary of $f_3$.
Since $\ell \geq |f_5| = 1$, we already know that we can at least extend phrase $f_5$ and thus set $p_1 := p = 3$ (line~\ref{algo:parse:extend}).
We do not need to call \text{LexSmallerPhrase} again as $3 = p_L \neq z-1 = 4$.
Then, because also $\ell \geq |f_5|+|f_4|=3$, we know that we can merge phrases $f_4$ and $f_5$ and thus set $p_2 := p = 3$ (line~\ref{algo:parse:merge}).

In the case distinction, we see that $p_2 = 3 \neq \bot$ and proceed to merge $f_4$ and $f_5$ to a phrase of length $|f_4|+|f_5|+1=4$, copying 3 characters from $f_{p_2} = f_3$ and appending the new character $S[7] = \texttt{\$}$.
As part of the merge, the association of position 5 in $\overleftarrow{S}$ as the boundary of phrase $f_4$ is removed from $M$.
\end{example}

\subsection{Analysis}
\label{sect:analysis}

Algorithm~\ref{algo:parse} is equivalent to the $\Oh{n \lg\lg n}$-time algorithm described by Kempa and Kosolobov~\cite{DBLP:conf/esa/KempaK17}.
We note that by using a balanced search tree for $M$, the time becomes $\Oh{n\lg z}$ as $M$ contains at most $\Th{z}$ entries.
This can be a useful alternative for large repetitive inputs.
The working space is clearly dominated by the array $A'$ and the LCP array $H$, i.e., $\Oh{n}$, also matching that of \cite{DBLP:conf/esa/KempaK17}.

Our practical improvements over \cite{DBLP:conf/esa/KempaK17} are threefold.
First, the associative nature of $M$ allows us to discard the suffix array $A$ after initialization.
Second, to ensure that $p_2 \neq z-1$, we only do one additional predecessor or successor query as needed, whereas~\cite{DBLP:conf/esa/KempaK17} temporarily removes the phrase $f_{z-1}$ from $M$.
This saves us a deletion from $M$ for every input character (and a re-insertion into $M$ in case we do not merge, see section~\ref{sect:repetitive-inputs} for details and side effects).
Third, we insert the previous phrase $f_z$ into $M$ \emph{lazily}, i.e., only once we begin a new phrase.
When trying to find an extension or merge candidate, similar to $p_2 \neq z-1$ when merging, it must hold that $p_1 \neq z$.
This is already guaranteed by doing the predecessor and succesor queries exluding $i'$ in \textsc{LexSmallerPhrase} and \textsc{LexGreaterPhrase}.
Therefore, it matters not if $i'$ has not yet been inserted into $M$.
Compared to \cite{DBLP:conf/esa/KempaK17}, where we have a guaranteed insertion into $M$ for every input character, the lazy insertion only occurs at most $\Th{z}$ times.
It also saves us the subsequent deletion of $f_z$ from $M$ when extended or merged.

\subsection{Preparing Efficient Random Access}

One of the key features of the LZ-End parsing is that it allows for efficient random access:
extracting a substring of length $\ell$ of the original input can be done in time $\Oh{\ell+h}$, where $h$ is the length of the longest LZ-End phrase in the parsing~\cite{DBLP:conf/dcc/KreftN10}.
It is therefore straightforward to make $h$ a parameter in the computation of the parsing to artificially limit the length of the longest phrase.
This offers a trade-off between compression and random access time.

Algorithm~\ref{algo:parse} can easily be modified to respect this:
in case $|f_z| = h$, we disallow any further extension of phrase $f_z$.
This check can be added to line~\ref{algo:parse:extend}.
Furthermore, if $|f_z| + |f_{z-1}| \geq h$, which we can check in line~\ref{algo:parse:merge}, we disallow merging the two.

\section{Implementation}
\label{sect:impl}

The pseudocode listing in Algorithm~\ref{algo:parse} is deliberately close to practical code.
Our C++ implementation is available at \url{https://github.com/pdinklag/lzend},
where we use \emph{libsais} for computing the suffix and LCP arrays, the data structure of \cite{DBLP:conf/latin/BenderF00} for range minimum queries, and a B-tree of \cite{DBLP:conf/wea/Dinklage0H21} as an ordered dictionary.

The associative predecessor and successor data structure $M$ could be implemented using any ordered dictionary (e.g., \texttt{std::map} or \texttt{java.util.TreeSet}).
Implementations of algorithms for computing the suffix and LCP array are widely available for many programming languages (e.g., through libraries \emph{libsais}\footnote{libsais: \url{https://github.com/IlyaGrebnov/libsais}} or \emph{jSuffixArrays}\footnote{jSuffixArrays: \url{http://labs.carrotsearch.com/jsuffixarrays.html}}).
A data structure for efficient (albeit not theoretically optimal) range minimum queries can be easily implemented by using a combination of blockwise sampling and scanning (as done, e.g., in \cite{DBLP:conf/esa/Dinklage0HKK20}).

To underline the practical relevancy of Algorithm~\ref{algo:parse}, we do a brief benchmark parsing inputs from the Pizza\&Chili Corpus\footnote{Pizza\&Chili Corpus: \url{http://pizzachili.dcc.uchile.cl/texts.html}}.
We compare our C++ implementation against the in-RAM implementation provided by Kempa in the \emph{lz-end-toolkit} mentioned in the beginning.
To the best of our knowledge, the latter is the current state of the art.
For a fair comparison, to account for the fact that different construction algorithms are used for the index data structures on the reversed input, we only measure the running time of the actual parsing phase.
Furthermore, for better performance, we set the integer width of the \emph{lz-end-toolkit} to four bytes.
The experiment was conducted on a Linux machine with an AMD~EPYC~7452 processor.

\begin{table}[t]
\begin{center}
\begin{tabular}{r | r r r | r r}
Input & $n$ & $z$ & $z/n$ & \emph{lz-end-toolkit} & Algorithm~\ref{algo:parse}\\
\hline
cere            & \num{461286644}  & \num{1863246}  & \num{0.40}\%  & \underline{\num{455}} & \num{582} \\
dblp.xml        & \num{296135874}  & \num{10244979} & \num{3.46}\%  & \num{283} & \underline{\num{251}} \\
dna             & \num{403927746}  & \num{26939573} & \num{6.67}\%  & \num{602} & \underline{\num{379}} \\
einstein.en.txt & \num{467626544}  & \num{104087}   & \num{0.02}\%  & \underline{\num{454}} & \num{728} \\
english.1024MB  & \num{1073741824} & \num{68034586} & \num{6.34}\%  & \num{2119} & \underline{\num{1160}} \\
pitches         & \num{55832855}   & \num{5675142}  & \num{10.16}\% & \num{27} & \underline{\num{25}} \\
proteins        & \num{1184051855} & \num{77369007} & \num{6.53}\%  & \num{1651} & \underline{1185} \\
sources         & \num{210866607}  & \num{12750341} & \num{6.05}\%  & \num{158} & \underline{135} \\
\end{tabular}
\end{center}
\caption{
    Benchmark results showing the parsing times, in seconds, of \emph{lz-end-toolkit} and Algorithm~\ref{algo:parse} (shortest underlined).
    For each input, we also list the length $n$, the number $z$ of LZ-End phrases and the ratio $z/n$ as a simple compressibility measure.
}
\label{tab:bench}
\end{table}

The results are shown in table~\ref{tab:bench}.
On most inputs, our implementation is much faster (up to nearly twice as fast on \emph{english.1024MB}) than the \emph{lz-end-toolkit}, indicating that the lazy evaluation of predecessor and successor queries as well as the removed layer of indirection via the suffix array $A$ can be very beneficial.

\subsection{Handling Highly Repetitive Inputs}
\label{sect:repetitive-inputs}

It stands out that the \emph{lz-end-toolkit} is faster than Algorithm~\ref{algo:parse} on highly repetitive inputs (namely \emph{cere} and \emph{einstein.en.txt}).
The reason is that it is implicitly tuned for this case.

Recall how, in contrast to \cite{DBLP:conf/esa/KempaK17}, we avoid the temporary removal of the boundary of phrase $f_{z-1}$ from $M$.
In their implementation (\emph{lz-end-toolkit}), they do the temporary removal and proceed to find a source phrase candidate for merging phrases $f_z$ and $f_{z-1}$.
If a merge is possible, they immediately do the merge and advance to the next input character.
The key difference to Algorithm~\ref{algo:parse} is that in this case,
(1) they do at most a predecessor and (only if no merge candidate is found) a successor query and
(2) the removal of $f_{z-1}$ from $M$, which we need to do as part of the merge, is already done.
However---and this is the reason why we chose to avoid a temporary modification of $M$---if merging is not possible, the boundary of phrase $f_{z-1}$ has to be re-inserted into $M$.

The conclusion is that if merges can be done frequently, the \emph{lz-end-toolkit} has to do less work than Algorithm~\ref{algo:parse}.
A practical observation is that for highly repetitive inputs, merges happen much more frequently compared to other less repetitive inputs.
However, Algorithm~\ref{algo:parse} has to do less work than the \emph{lz-end-toolkit} in the case that no merge is possible.
These facts explain the observed discrepancies in the running time comparison.

Our takeaway is that Algorithm~\ref{algo:parse} can easily be tuned for highly repetitive inputs by re-arranging the candidate search to first look for merges, including a preliminary removal of $f_{z-1}$ from $M$ to take advantage of.

\clearpage
\bibliography{literature}

\end{document}